\documentstyle[12pt]{article}
\newcommand{\EQ}{\begin{equation}}
\newcommand{\EN}{\end{equation}}
\newcommand{\bea}{\begin{eqnarray}}
\newcommand{\eea}{\end{eqnarray}}
\newcommand{\hs}{\hspace{0.1cm}}

\newcommand{\goto}{\rightarrow}

\begin{document}
\setcounter{page}{0}
\topmargin 0pt
\oddsidemargin 5mm
\renewcommand{\thefootnote}{\arabic{footnote}}
\newpage
\setcounter{page}{0}
\begin{titlepage}
\begin{flushright}
\end{flushright}
\vspace{0.5cm}
\begin{center}
{\large {\bf Universal amplitude ratios in the two-dimensional Ising 
model\footnote{Work supported by the European Union under contract 
FMRX-CT96-0012}}}\\
\vspace{1.8cm}
{\large Gesualdo Delfino} \\
\vspace{0.5cm}
{\em Laboratoire de Physique Th\'eorique, Universit\'e de Montpellier II}\\
{\em Pl. E. Bataillon, 34095 Montpellier, France}\\
{\em E-mail: aldo@lpm.univ-montp2.fr}\\
\end{center}
\vspace{1.2cm}

\renewcommand{\thefootnote}{\arabic{footnote}}
\setcounter{footnote}{0}

\begin{abstract}
\noindent
We use the results of integrable field theory to determine the universal 
amplitude ratios in the two-dimensional Ising model. In particular, the
exact values of the ratios involving amplitudes computed at nonzero 
magnetic field are provided.
\end{abstract}

\vspace{.3cm}

\end{titlepage}

\newpage
Universality is one of the most fascinating concepts of statistical mechanics
\cite{Cardy}.
Briefly stated, it says that physical systems with different microscopic 
structure but having in common some basic internal symmetry exhibit the 
same critical behaviour in the vicinity of a phase transition point. The 
point is best illustrated considering the singular behaviour of the various 
thermodynamic quantities nearby the critical point. For a magnetic system 
exhibiting a second order phase transition the usual notation is 
\bea
& C\simeq (A/\alpha)\,\tau^{-\alpha}\,, & \hspace{1cm}\tau>0\,,
\hspace{.2cm}h=0 \nonumber\\
& C\simeq (A'/\alpha')\,(-\tau)^{-\alpha'}\,, & \hspace{1cm}\tau<0\,,
\hspace{.2cm}h=0 \nonumber\\
& C\simeq (A_c/\alpha_c)\,|h|^{-\alpha_c}\,, & \hspace{1cm}\tau=0\,,
\hspace{.2cm}h\neq 0 \nonumber\\
& M\simeq B\,(-\tau)^{\beta}\,, & \hspace{1cm}\tau<0\,,
\hspace{.2cm}h=0^+ \nonumber\\
& \chi\simeq\Gamma\,\tau^{-\gamma}\,, & \hspace{1cm}\tau>0\,,\hspace{.2cm}h=0
\nonumber\\
& \chi\simeq\Gamma'\,(-\tau)^{-\gamma'}\,, & \hspace{1cm}\tau<0\,,
\hspace{.2cm}h=0 \nonumber\\
& \chi\simeq\Gamma_c\,|h|^{-\gamma_c}\,, & \hspace{1cm}\tau=0\,,
\hspace{.2cm}h\neq 0 \nonumber\\
& h\simeq D_c\,M|M|^{\delta-1}\,, & \hspace{1cm}\tau=0\,,
\hspace{.2cm}h\neq 0 \nonumber\\
& \xi\simeq\xi_0\,\tau^{-\nu}\,, & \hspace{1cm}\tau>0\,,
\hspace{.2cm}h=0 \nonumber\\
& \xi\simeq \xi_0'\,(-\tau)^{-\nu'}\,, & \hspace{1cm}\tau<0\,,
\hspace{.2cm}h=0 \nonumber\\
& \xi\simeq\xi_c\,|h|^{-\nu_c}\,, & \hspace{1cm}\tau=0\,,
\hspace{.2cm}h\neq 0 \nonumber
\eea
where $\tau=a\,(T-T_c)$ ($a$ positive constant), $h$ is the applied magnetic
field and the limit towards the critical point $\tau=0$, $h=0$ is understood.
$C$, $M$, $\chi$ and $\xi$ denote the specific heat, the 
magnetisation, the susceptibility and the correlation length, 
respectively.

The critical exponents $\alpha$, $\beta$,.. are the same for all systems 
within a given universality class and are related by the scaling and 
hyperscaling relations in $d$ dimensions
\bea
&& \alpha=\alpha'\,,\hspace{.4cm}\gamma=\gamma'\,,\hspace{.4cm}\nu=\nu'\,,
\nonumber\\
&& \gamma=\beta(\delta-1)\,,\hspace{.4cm}\alpha=2-2\beta-\gamma\,,
\hspace{.4cm}2-\alpha=d\nu\,, \nonumber\\
&& \alpha_c=\alpha/\beta\delta\,,\hspace{.4cm}\gamma_c=1-1/\delta\,,
\hspace{.4cm}\nu_c=\nu/\beta\delta\,. \nonumber
\eea
The critical amplitudes $A$, $B$,.., on the other hand, depend on the scale 
factors used for $\tau$ and $h$ and are nonuniversal. However, universal
ratios of amplitudes can be constructed in which any dependence on metric 
factors cancels out. Together with the critical exponents, these ratios 
further characterise the given universality class. The standard amplitude 
combinations considered in the literature are \cite{PHA}
\bea
&& A/A'\,,\hspace{.5cm}\Gamma/\Gamma'\,,\hspace{.5cm}\xi_0/\xi_0'\,, 
\label{tr1}\\
&& R_C=A\Gamma/B^2\,,\hspace{.5cm}R_\xi^+=A^{1/d}\xi_0\,,
\label{tr2}\\
&& R_\chi=\Gamma D_c B^{\delta-1}\,,\hspace{.5cm}
R_A=A_cD_c^{-(1+\alpha_c)}B^{-2/\beta}\,,\hspace{.5cm}
Q_2=(\Gamma/\Gamma_c)(\xi_c/\xi_0)^{\gamma/\nu}\,.
\label{mr}
\eea

\vspace{.3cm}
A substantial progress was made over the last years in the derivation of 
nonperturbative theoretical results in two-dimensional statistical mechanics 
and quantum field theory. The solution of conformal field theories (CFTs) 
\cite{BPZ,ISZ} provided an almost complete classification of universality 
classes for second order phase transitions in $d=2$. In particular, it solved 
the problem of the exact determination of the critical exponents. The critical
amplitudes, however, carry information about the scaling region outside the 
critical point and are not determined by CFT. In this respect, the 
Zamolodchikov's observation that specific perturbations of the critical point 
lead to integrable off-critical theories \cite{Sacha} is of crucial 
importance. The integrable theories obtained in this way, regarded as 
quantum field theories in $1+1$ dimensions, are characterisable through
the determination of their exact $S$-matrix. A number of physical quantities 
can then be computed using different techniques \cite{report}. In particular,
the results provided by the thermodynamyc Bethe ansatz (TBA) \cite{TBA} and 
the form factor approach \cite{KW,Smirnov,YZ,CM1} enable 
the determination of amplitude ratios and have been used for this purpose
in the problems of self-avoiding walks \cite{CM2} and percolation \cite{CD}
(see also \cite{Smilga}).

It is the purpose of this note to illustrate the derivation of universal
amplitude ratios from (integrable) field theory through the very basic example 
of the two-dimensional Ising model. Of course, the purely ``thermal'' ratios 
(\ref{tr1}) and
(\ref{tr2}) are exactly known for this case since the seventies, when the 
correlation functions of the Ising model at $h=0$ were first computed on the 
lattice \cite{Mccoy-Wu,WMcTB}. The possibility to determine the ratios 
(\ref{mr}), on the contrary, relies on the more recent realisation that
the scaling limit of the Ising model at $\tau=0$ and $h\neq 0$ is an 
integrable theory \cite{Sacha}.

We will regard the scaling limit of the two-dimensional Ising model as 
described by the euclidean field theory defined by the action
\EQ
{\cal S}={\cal S}_{CFT}-\tau\int d^2x\,\varepsilon(x)-
h\int d^2x\,\sigma(x)\,\,,
\label{action}
\EN
where ${\cal S}_{CFT}$ stays for the Ising critical point conformal action
and $\sigma$ and $\varepsilon$  denote the magnetisation and energy operators,
respectively. All the critical exponents are determined by the scaling 
dimensions $x_\sigma=1/8$ and $x_\varepsilon=1$ of the operators $\sigma$ 
and $\varepsilon$: $\alpha=(d-2x_\varepsilon)/(d-x_\varepsilon)=0$, 
$\beta=x_\sigma/(d-x_\varepsilon)=1/8$, $\gamma=7/4$, $\delta=15$, $\nu=1$, 
$\alpha_c=0$, $\gamma_c=14/15$, $\nu_c=8/15$. 
The physical dimensions of the two couplings in (\ref{action}) are
$\tau\sim m^{2-x_\varepsilon}$ and $h\sim m^{2-x_\sigma}$, $m$ being a mass
parameter. What is important for us is that the theory (\ref{action}) becomes
integrable when at least one of the two couplings is set to zero \cite{Sacha}.

Choosing the scale factors for $\tau$ and $h$ amounts to fixing a
normalisation for the conjugated operators $\varepsilon$ and $\sigma$. We 
will proceed to determine the critical amplitudes within the standard CFT 
normalisation defined by the asymptotic conditions
\bea
&& \langle\sigma(x)\sigma(0)\rangle=|x|^{-1/4}\,,\hspace{.5cm}x\goto 0
\label{sigmanorm} \nonumber\\
&& \langle\varepsilon(x)\varepsilon(0)\rangle=|x|^{-2}\,,
\hspace{.5cm}x\goto 0\,\,.
\label{epsnorm}
\eea

Let us begin with the amplitudes for the correlation length. The latter can
be defined in different ways. For the time being, 
we will refer to the so called ``true'' correlation length defined through 
the large distance decay of the spin-spin correlation function
\EQ
\langle\sigma(x)\sigma(0)\rangle^c\sim \frac{e^{-|x|/\xi}}{|x|^{(d-1)/2}}\,,
\hspace{.5cm}|x|\goto\infty 
\EN
where $\langle\cdots\rangle^c$ denotes the connected correlator. From the 
representation of the correlator as a spectral sum over $n$-particle 
intermediate states
\EQ
\langle\sigma(x)\sigma(0)\rangle^c=\sum_n|\langle 0|\sigma(0)|n\rangle|^2
e^{-E_n|x|}\,,
\label{spectral}
\EN
it is clear that the ``true'' correlation length is the inverse of the total 
mass of the lightest state entering the decomposition above. For integrable
models, the exact relation between the particle masses and the couplings 
appearing in the action is provided by the TBA \cite{sgmass-Fateev}. 
The result
for the Ising model is\footnote{We denote by the subscript $\tau$ ($h$) the 
quantities referring to the theory (\ref{action}) with $h=0$ ($\tau=0$). 
$m_h$ is the mass of the lightest among the $8$ particles the mass spectrum of 
the $\tau=0$ theory consists of.}
\bea
&& m_\tau=2\pi\,|\tau|\,, \nonumber\\
&& m_h={\cal C}\,|h|^{8/15}\,, \nonumber\\
&& {\cal C}=\frac{4 \sin\frac{\pi}{5} \Gamma(1/5)}{\Gamma(2/3)\Gamma(8/15)}
\left(\frac{4\pi^2\Gamma(3/4)\Gamma^2(13/16)}{\Gamma(1/4)\Gamma^2(3/16)}
\right)^{4/15}=4.40490857..\,\,. 
\eea
Due to the invariance under spin reversal at $h=0$, the magnetisation operator
couples only to the states with odd (even) number of particles for $\tau>0$
($\tau<0$).When $h\neq 0$, $\sigma$ couples to any intermediate state.
It follows
\bea
&&\xi_0=2\xi_0'=1/(2\pi)\,, \nonumber\\
&&\xi_c=1/{\cal C}\,\,.
\eea

The specific heat diverges logarithmically in the Ising model ($\alpha=0$) and
the specific heat amplitudes are accordingly defined through 
$C\simeq-A\ln\tau$ and analogous relations for $A'$ and $A_c$.
Writing the partition function as $\mbox{Tr}\,\exp(-{\cal S})$ and the reduced 
free energy as $f=-1/V\ln Z$, the specific heat per unit volume is given by
\EQ
C=-\frac{\partial^2f}{\partial\tau^2}=\int d^2x\,\langle
\varepsilon(x)\varepsilon(0)\rangle^c\sim-2\pi\ln mr_0\,, 
\EN
where (\ref{epsnorm}) was used and $r_0$ is an ultraviolet cutoff. 
Recalling the relations between the mass scale $m$ and the coupling constants
$\tau$ and $h$, the specific heat amplitudes in the 
CFT normalisation are simply
\EQ
A=A'=2\pi\,,\hspace{1.5cm}A_c=2\pi\,\frac{8}{15}.
\EN

The magnetisation per unit volume is given by
\EQ
M=-\frac{\partial f}{\partial h}=\langle\sigma\rangle\,\,. 
\EN
Vacuum expectation values in the CFT normalisation are exactly known for 
integrable models due to the TBA \cite{sgmass-Fateev} and some more recent
developments \cite{FLZZ}. For the magnetisation operator in the Ising model
at $\tau<0,h=0$ and at $\tau=0,h>0$ they are, respectively\footnote{In the 
following we will consider for convenience only positive values of $h$. Due to
the simmetry about $h=0$ this involves no loss of generality.}
\bea
&&\langle\sigma\rangle_\tau=B\,(-\tau)^{1/8}\,, \nonumber\\
&&\langle\sigma\rangle_h=\frac{2\,{\cal C}^2}{15\,(\sin\frac{2\pi}{3}+
\sin\frac{2\pi}{5}+\sin\frac{\pi}{15})}\,h^{1/15}=1.27758227..\,h^{1/15}\,,
\label{sigmah}
\eea
with
\EQ
B=2^{1/12}(2\pi/e)^{1/8}{\cal A}^{3/2}=1.70852190..\,\,;
\EN
we used the Glaisher's constant
\EQ
{\cal A}=2^{7/36}\pi^{-1/6}\exp\left[\frac{1}{3}+\frac{2}{3}
\int_0^{1/2}dx\ln\Gamma(1+x)\right]=1.28242712..\,\,.
\EN
From (\ref{sigmah}) we deduce
\EQ
D_c=\left[\frac{15\,(\sin\frac{2\pi}{3}+
\sin\frac{2\pi}{5}+\sin\frac{\pi}{15})}{2\,{\cal C}^2}\right]^{15}
=0.0253610264..\,\,.
\EN

The reduced susceptibility is defined as
\EQ
\chi=-\frac{\partial^2f}{\partial h^2}=\int d^2x\,\langle
\sigma(x)\sigma(0)\rangle^c\,\,. 
\EN
The amplitudes $\Gamma$ and $\Gamma'$ were computed exactly in \cite{WMcTB}
integrating the $h=0$ spin-spin correlator. In the CFT normalisation they 
read\footnote{The relations one needs to pass from the lattice normalisation
of \cite{WMcTB} to the one used here are $\sigma_{lat}=2^{5/48}e^{1/8}{\cal A}
^{-3/2}\,\sigma$ and $[(T-T_c)/T_c]_{lat}=(\pi/\ln(1+\sqrt 2))\,\tau$.}
\bea
&& \Gamma=0.148001214..  \,, \nonumber\\
&& \Gamma'=0.00392642280.. \,\,.
\eea
When $\tau=0$, $\sigma$ is the operator which perturbs the critical point. 
The zeroth moment of the spin-spin correlator is then related to the vacuum
expectation value as
\EQ
\int d^2x\,\langle\sigma(x)\sigma(0)\rangle^c_h=\frac{x_\sigma}{(2-x_\sigma)h}
\,\langle\sigma\rangle_h\,\,.
\label{zeroth}
\EN
It follows
\EQ
\Gamma_c=\frac{1}{\delta}D_c^{-1/\delta}=0.0851721517..  \,\,.
\EN

\vspace{.3cm}
The dependence on the operator normalisations (the only nonuniversal 
ingredient we had in our computation) cancels when the combinations
(\ref{tr1}), (\ref{tr2}) and (\ref{mr}) are considered. The exact results we
obtain for the ratios are collected in Table\,1. The values for the
combinations of purely thermal amplitudes are well known. Concerning the 
ratios involving amplitudes computed at $h\neq 0$, to the best of our knowledge
the only reliable extimates in $d=2$ come from the series analysis of 
Ref.\,\cite{TF}. The value $R_\chi\simeq 6.78$ is quoted there together with a 
value for $Q_2$ which uses the ``second moment'' correlation length (see 
below). No previous result for $R_A$ is known to us.

We conclude this note computing the ratios which involve the correlation length
amplitudes using the second moment correlation length
\EQ
\xi^2_{2nd}\equiv\frac{1}{2d}\,\frac{\int d^2x\,|x|^2\langle\sigma(x)\sigma(0)
\rangle^c}{\int d^2x\,\langle\sigma(x)\sigma(0)\rangle^c}\,\,. 
\EN
The amplitudes $(\xi_0)_{2nd}$ $(\xi_0')_{2nd}$  could be exactly computed 
by numerical integration of the spin-spin correlation function at $h=0$. 
Here, we will take a short cut with the purpose of illustrating a point
which is particularly relevant for more general applications. In integrable 
models the spectral sum (\ref{spectral}) is not
a purely formal expression because the matrix elements it contains (known as 
form factors) can be computed exactly. While the Ising model at $h=0$
remains the single example for which the resummation of the spectral series is 
known, a remarkably fast convergence of the series emerged in the last years 
as a general feature of integrable models \cite{YZ,CM2,immf}. This important 
circumstance makes the form factor approach extremely effective for obtaining
accurate quantitative results at a relatively little cost.
 
We compute $(\xi_0)_{2nd}$ and
$(\xi_0')_{2nd}$ in the leading form factor approximation ($1$-particle 
contribution for $\tau>0$, $2$-particle contribution for $\tau<0$). 
Using\footnote{On-shell momenta are parameterised as $p^\mu=(m_\tau\cosh\theta,
m_\tau\sinh\theta)$.} \cite{BKW}
\bea
&&\langle 0|\sigma(0)|\theta\rangle_\tau=\langle\sigma\rangle_\tau\,,
\nonumber\\
&&\langle 0|\sigma(0)|\theta_1,\theta_2\rangle_\tau=
i\langle\sigma\rangle_\tau\,\tanh\frac{\theta_1-\theta_2}{2}\,, 
\eea
one finds
\bea
&&(\xi_0)_{2nd}\simeq 1/(2\pi)\,, \nonumber\\
&&(\xi_0')_{2nd}\simeq 1/(2\sqrt{10}\,\pi)\,\,.
\eea
$(\xi_c)_{2nd}$ could be similarly extimated using the form factors for the 
$\tau=0$ model computed in \cite{immf}. However, its exact value is 
immediately determined reminding that $\sigma$ is the perturbing operator
at $\tau=0$. Then the $c$-theorem sum rule holds \cite{cth}
\EQ
c=\frac{3}{4\pi}[2\pi h(2-x_\sigma)]^2\int d^2x\,|x|^2\langle\sigma(x)
\sigma(0)\rangle^c_h\,, 
\EN
where the central charge $c$ equals $1/2$ for the Ising universality class.
Combining with (\ref{zeroth}) one gets
\EQ
(\xi_c)_{2nd}=\sqrt{\frac{8}{45\pi}}D_c^{1/30}=0.21045990..\,\,.
\EN
Table\,2 contains the results one obtains for the amplitudes ratios. The 
value $(\xi_0/\xi_0')_{2nd}=3.16..$ is quoted in Ref.\,\cite{Bartelt} as the 
result of the integration of the exact correlator; $(Q_2)_{2nd}=2.88\pm 0.02$ 
is the series result obtained in Ref.\,\cite{TF}.


\newpage
\hs
\vspace{25mm}

{\bf Table Caption}

\vspace{1cm}
\begin{description}
\item [Table 1]. Exact amplitude ratios for the two-dimensional Ising model.
The results for $\xi_0/\xi_0'$, $R_\xi^+$ and $Q_2$ refer to the ``true''
correlation length.
\item [Table 2]. Results referring to the ``second moment'' correlation
length with $(\xi_0)_{2nd}$ and $(\xi_0')_{2nd}$ computed in the leading form
factor approximation.

\end{description}

\newpage

\begin{center}

\vspace{3cm}
\begin{tabular}{|c|}\hline
$ A/A' = 1 $  \\
$ \Gamma/\Gamma' = 37.6936520.. $ \\
$ \xi_0/\xi_0' = 2 $ \\
$ R_C = 0.318569391.. $  \\
$ R_\xi^+ = 1/\sqrt{2\pi} $  \\
$ R_\chi = 6.77828502.. $  \\
$ R_A = 0.0250658794.. $  \\
$ Q_2 = 3.23513834.. $  \\ \hline
\end{tabular}
\end{center}

\begin{center}
{\bf Table 1}
\end{center}

\begin{center}

\vspace{3cm}
\begin{tabular}{|c|}\hline
$ (\xi_0/\xi_0')_{2nd}\simeq 3.162 $ \\
$ (R_\xi^+)_{2nd}\simeq 0.3989 $  \\
$ (Q_2)_{2nd}\simeq 2.833 $  \\ \hline
\end{tabular}
\end{center}

\begin{center}
{\bf Table 2}
\end{center}

\end{document}